\documentclass[graybox]{svmult}

\usepackage{mathptmx}       
\usepackage{helvet}         
\usepackage{courier}        
\usepackage{type1cm}        
\usepackage{makeidx}         
\usepackage{graphicx}        
\usepackage{multicol}        
\usepackage[bottom]{footmisc}

\makeindex             

\begin{document}

\title*{Magnetoelectric effects, helical phases, and FFLO phases}
\author{D.F. Agterberg}
\institute{D.F. Agterberg \at Department of Physics, University of Wisconsin-Milwaukee, Milwaukee, WI 53211, US, \email{agterber@uwm.edu}}
%
%
\maketitle

\abstract*{This chapter emphasizes new magnetic properties that arise when inversion
symmetry is broken in a superconductor. There are two aspects that will be
covered in detail.
The first topic encompasses physics related to superconducting
magnetoelectric effects that arise from broken inversion symmetry. Broken
inversion symmetry allow for Lifshitz invariants in the free energy which can be
viewed as a coupling between the magnetic induction and the supercurrent.
There are similarities between these invariants and the better known
Dzyaloshinskii-Moyira interaction in magnetic systems. These Lifshitz invariants
give rise to anomalous magnetic properties as well as new phases in the
presence of magnetic fields. Here, we will describe the consequences of these
Lifshitz invariants, provide estimates for the relative magnitudes of the novel
effects, and discuss the important role that crystal symmetry plays in
understanding this physics.
Finally, we provide a discussion of the fate of Fulde-Ferrell-Larkin-Ovchinnikov
(FFLO) phases in broken inversion superconductors. In particular, we show how
broken inversion symmetry can have a profound effect on the stability, existence,
and properties of FFLO phases.}

\abstract{This chapter emphasizes new magnetic properties that arise when inversion
symmetry is broken in a superconductor. There are two aspects that will be
covered in detail.
The first topic encompasses physics related to superconducting
magnetoelectric effects that arise from broken inversion symmetry. Broken
inversion symmetry allow for Lifshitz invariants in the free energy which can be
viewed as a coupling between the magnetic induction and the supercurrent.
There are similarities between these invariants and the better known
Dzyaloshinskii-Moyira interaction in magnetic systems. These Lifshitz invariants
give rise to anomalous magnetic properties as well as new phases in the
presence of magnetic fields. Here, we will describe the consequences of these
Lifshitz invariants, provide estimates for the relative magnitudes of the novel
effects, and discuss the important role that crystal symmetry plays in
understanding this physics.
Finally, we provide a discussion of the fate of Fulde-Ferrell-Larkin-Ovchinnikov
(FFLO) phases in broken inversion superconductors. In particular, we show how
broken inversion symmetry can have a profound effect on the stability, existence,
and properties of FFLO phases.}

\section{Introduction}
\label{sec:1}
One important way in which non-centrosymmetric superconductors differ from conventional superconductors is in the response to magnetic fields. In particular, the removal of inversion symmetry leads to new terms in the free energy that give rise to magneto-electric effects. These effects are closely related to the appearance of magnetic field generated helical phase in which the superconducting order develops a periodic spatial variation.  Here we review this physics beginning with a detailed examination of the phenomenological theory followed by an overview of microscopic treatments of these problems which include an overview an of the interplay of the helical phase and Fulde-Ferrell-Larkin-Ovchinnikov (FFLO) phases \cite{ff,lo}.

\section{Phenomenology of Single Component Superconductors}
\label{sec:2}

This section reviews the phenomenology relating Lifshitz invariants in the the free energy to magnetoelectric effects, vortex structures, and the helical phase.

\subsection{Ginzburg Landau free energy}
\label{subsec:2}
A key new feature of non-centrosymmetric superconductors is the existence of Lifshitz invariants in the Ginzburg Landau (GL) free energy \cite{MS94,Edel96,Y02,agt03,KAS05,MinSam08}. These give rise to magnetoelectric effects \cite{LNE85,Edel95,Y02,Fuj05,LuY08,LuY09}, helical phases \cite{agt03,DF03,KAS05,dim07,agt07}, and novel magnetic properties \cite{LNE85,KAS05,OIM06,yip07,yip-cond-mat,LuY08} discussed in this chapter. To examine the consequences of these invariants we initially consider a GL theory for a single component order parameter (for example, an $s$-wave superconductor) and add the most general Lifshitz invariant allowed by broken inversion symmetry. Specific Lifshitz invariants are tabulated in Table 1 for different point group symmetries of the material in question. Since the primary goal is to reveal the new physics arising from these invariants, we ignore the role of any anisotropy that might appear in the usual GL free energy. Under these conditions the GL free energy under consideration is (we work in units such that $\hbar=c=1$):
\begin{equation}
F=\int d^3r\left\{\alpha|\eta|^2
+K\eta^*{\bf D}^2\eta
+ K_{ij}B_i[\eta^*(D_j\eta)+\eta(D_j\eta)^*] +\frac{\beta}{2}|\eta|^4+\frac{B^2}{8\pi}\right\}, \label{glfree}
\end{equation}
where $\alpha=\alpha_0(T-T_c)$, $D_i=-i\nabla_i-2eA_i$ and ${\bf B}=\nabla\times {\bf A}$.
From this free energy, the GL equations can be found by varying the above with respect to ${\bf A}$ and $\eta$. This results in the following:
\begin{equation}
\alpha\eta+\beta|\eta|^2\eta+K{\bf D}^2\eta+K_{ij}[2h_i(D_j\eta)+i\eta\nabla_jB_i]=0
\end{equation}
and
\begin{equation}
{\bf J}_i=\frac{1}{4\pi}[\nabla\times({\bf B}-4\pi {\bf M})]_i=2eK[\eta^*(D_i\eta)+\eta(D_i\eta)^*]+4eK_{ji}B_j|\eta|^2
\label{GLcurrent}
\end{equation}
where
\begin{equation}
{\bf M}_i=-K_{ij}[\eta^*(D_j\eta)+\eta(D_j\eta)^*] \label{mag1}.
\end{equation}
These equations are joined by the boundary conditions (which follow from the surface terms that arise from integration by parts in the variation of $F$):
\begin{equation}
[K\hat{n}_i(D_i\eta)+K_{ij}B_i\hat{n}_j\eta]_{boundary}=0 \label{boundary}
\end{equation}
where $\hat{n}_j$ is the component of the surface normal along $\hat{j}$, and the usual Maxwell boundary conditions on the continuity of the normal component of ${\bf B}$ and the transverse components of  ${\bf H}={\bf B}-4\pi {\bf M}$ (the appearance of ${\bf M}$ due to the Lifschitz invariants makes this boundary condition non-trivial). Note that adding the complex conjugate of Eq.~\ref{boundary} multiplied by $\eta^*$ to Eq.~\ref{boundary} multiplied by $\eta$  yields ${\bf J}\cdot {\hat n}|_{boundary}=0$.

\begin{table}
\begin{tabular}{|c|c|}
\hline
Point Group & Lifshitz Invariants\\
  \hline
  $O$ & $K(B_xj_x+B_yj_y+B_zj_z)$ \\
  $T$ &  $K(B_xj_x+B_yj_y+B_zj_z)$\\
  $D_6$ & $K_1(B_xj_x+B_yj_y+B_zj_z)+K_2B_zj_z$ \\
  $C_{6v}$ & $K(B_xj_y-B_yj_x)$ \\
  $C_6$ & $K_1(B_xj_x+B_yj_y+B_zj_z)+K_2B_zj_z+K_3(B_xj_y-B_yj_x)$ \\
  $D_4$ & $K_1(B_xj_x+B_yj_y+B_zj_z)+K_2B_zj_z$ \\
  $C_{4v}$ & $K(B_xj_y-B_yj_x)$ \\
  $D_{2d}$ & $K(B_xj_y-B_yj_x)$ \\
  $C_4$ & $K_1(B_xj_x+B_yj_y+B_zj_z)+K_2B_zj_z+K_3(B_xj_y-B_yj_x)$  \\
  $S_4$ & $K_1(B_xj_x-B_yj_y)+K_2(B_yj_x+B_xj_y)$  \\
  $D_3$ & $K_1(B_xj_x+B_yj_y+B_zj_z)+K_2B_zj_z$ \\
  $C_{3v}$ & $K(B_xj_y-B_yj_x)$ \\
  $C_3$ & $K_1(B_xj_x+B_yj_y+B_zj_z)+K_2B_zj_z+K_3(B_xj_y-B_yj_x)$ \\
  $D_2$ & $K_1B_xj_x+K_2B_yj_y+K_3B_zj_z$ \\
  $C_{2v}$ & $K_1B_xj_y+K_2B_yj_x$ \\
  $C_2$ & $K_1B_xj_x+K_2B_yk_y+K_3B_zj_z+K_4B_yj_x+K_5B_xj_y$ \\
  $C_s$ & $K_1B_zk_x+K_2B_zj_j+K_3B_xj_z+K_4B_yj_z$\\
  $C_1$ & all components allowed \\
  \hline
\end{tabular}
\caption{Allowed Lifshitz invariants for different point groups. Here $j_i=\eta^*(D_i\eta)+\eta(D_i\eta)^*$.}
\end{table}

The appearance of ${\bf M}$ in Eq.~\ref{mag1} and the associated magnetization current leads to new physics in non-centrosymmetric superconductors.  Also note, as is the case for centrosymmetric superconductors, the boundary conditions are valid on a length scale greater that $\xi_0$, the zero-temperature coherence length. In the following few subsections,
 we present the solution to some common problems to provide insight into the role of the Lifshitz invariants.

\subsection{Solution with a spatially uniform Magnetic field: Helical Phase}

\label{subsec:3}

In situations when the magnetic field is spatially uniform, the GL equations describing the physics can be greatly simplified by introducing the following new order parameter:
\begin{equation}
\tilde{\eta}=\eta \exp\big(i{\bf q}{\cdot {\bf x}}\big)=\eta \exp\big(i\frac{iB_jK_{jk}x_k}{K}\big). \label{hel}
\end{equation}
The GL free energy for $\tilde{\eta}$ no longer has any Lifshitz invariants and is
\begin{equation}
F=\int d^3r\left\{\Big[\alpha-B_lK_{lm}B_jK_{jm}\Big]|\tilde{\eta}|^2
+K_1\tilde{\eta}^*{\bf D}^2\tilde{\eta}
 +\frac{\beta}{2}|\tilde{\eta}|^4+\frac{B^2}{8\pi}\right\}. \label{newGL}
\end{equation}
The resulting new GL equations are now those of a single component superconductor with a magnetic field induced enhancement of $T_c$ (this magnetic field enhancement is discussed in more detail in Chapter 1).  These new GL equations follow from a minimization of Eq.~\ref{newGL} with respect to ${\bf A}$ and $\tilde{\eta}$. Note that the phase factor introduced above cancels the additional current contribution from the Lifshitz invariants in Eq.~\ref{GLcurrent} and also cancels the related Lifshitz invariant contribution to the boundary condition. Furthermore, the magnetization that follows from Eq.~\ref{newGL} by taking the derivative with respect to $B_i$ coincides with that due Eq.~\ref{mag1} found prior to the redefinition of the order parameter. This modified free energy of Eq.~\ref{newGL} immediately implies that some results from the usual GL theory apply. In particular:\\

\noindent i) the vortex lattice solution near the upper critical field is the same as that of Abrikosov.\\
ii) the surface critical field $H_{c3}$ is the same as that of DeGennes. The order $B^2$ corrections to $T_c$ do not change $H_{c3}$ to leading order in $(T_c-T)/T_c$.\\
iii) the critical current in this wires will show no unusual asymmetry (this conclusion differs from that of Ref.~\cite{Edel96}).\\

\subsubsection{Helical Phase}

The main new feature that appears in a uniform magnetic field is the spatial modulation of the order parameter. Since $\eta$ develops a helical spatial dependence in the complex plane, the resulting thermodynamic phase has been named the helical phase. Since helicity of the order parameter is related to
its phase, an interference experiment based on the Josephson
effect would provide the most reliable test to observe this. Indeed, such an experiment has been proposed \cite{KAS05}. In particular, consider the example of a 2D non-centrosymmetric superconductor (with a Rashba spin-orbit interaction) with a Zeeman field applied in the 2D plane. Then consider a Josephson junction between this and another thin film
superconductor that is centrosymmetric. For a magnetic field applied in the
plane of the film {\it perpendicular} to the junction and with the non-centrosymmetric
superconductor oriented so that the helicity ${\bf q}$ is
perpendicular to the field ; we find this gives rise to an
interference effect analogous to the standard Fraunhofer pattern.
For this experiment, the film must be sufficiently thin so that the
magnetic field and the magnitude of the order parameter are
spatially uniform.

To illustrate this, consider the following free energy of the
junction
\begin{equation}
H_J=-t\int dx[\Psi_1({\bf x})\Psi_2^*({\bf x})+c.c.]
\end{equation}
where the integral is along the junction. The resulting Josephson
current is
\begin{equation}
I_J=Im\Big [ t\int dx\Psi_1({\bf x})\Psi_2^*({\bf x})\Big ]
\end{equation}
Setting the junction length equal to $2L$, and integrating yields
a maximum Josephson current of
\begin{equation}
I_J=2t|\Psi_{1}^0||\Psi_{2}^0|\frac{|\sin(qL)|}{|q L|}
\end{equation}
This demonstrates that the Josephson current will
display an interference pattern for a field {\it perpendicular} to
the junction. Note that in the usual case the Fraunhofer pattern
would be observed for a magnetic field perpendicular to the thin
film for which a finite flux passes through the junction.

\subsubsection{Magnetoelectric Effect}

Amongst the early theoretical studies of non-centrosymmetric superconductors, it was pointed out that a supercurrent must be accompanied by a spin polarization of the carriers \cite{Edel95}. Within the macroscopic theory given above, this spin polarization is described by the magnetization in Eq.~\ref{mag1}.  This magnetization appears when the supercurrent is non-vanishing due to a finite phase gradient. Subsequent to this proposal, it was suggested that the converse effect would also appear: a Zeeman field would induce a supercurrent \cite{Y02}. This would follow from the expression for the current of Eq.~\ref{GLcurrent} when the usual GL current ($2eK[\eta^*(D_i\eta)+\eta(D_i\eta)^*]$) vanishes. However, the latter proposal does not include the possibility discussed above that the order parameter develops a spatial modulation in the presence of a spatially homogeneous magnetic field (which leads to a nonvanishing $2eK[\eta^*(D_i\eta)+\eta(D_i\eta)^*]$). Indeed, this new equilibrium state ensures that the resultant supercurrent is vanishing. Nevertheless, as pointed out in Ref.~\cite{DF03}, it is possible to create this current using a geometry similar to that used to observe Little-Parks oscillations. In particular, the supercurrent has two contributions, one is the current due to the Lifshitz invariants and the other is the usual GL current $2eK[\eta^*(D_i\eta)+\eta(D_i\eta)^*]$. In the helical phase, these two contributions exactly cancel. By wrapping the superconductor in a cylinder, the condition that the order parameter is single valued does not allow the helical phase to fully develop since arbitrary spatial oscillations are not allowed. Consequently, when a magnetic field is applied along the cylindrical axis, a non-zero current can flow.  The resulting current will develop a periodic dependence on the applied magnetic field \cite{DF03}.

\subsection{London Theory and Meissner State}
\label{subsec:4}
We now turn to situations in which the magnetic field is not spatially uniform. The Lifshitz invariants lead to new physics for both the single vortex solution and for the usual penetration depth problem. To see this, we begin with the London limit and set $\eta=|\eta|e^{i\theta}$ and assume that the magnitude $|\eta|$ is fixed.  The GL free energy is then minimized with respect to $\theta$ and ${\bf A}$. The minimization with respect to $\theta$ yields
\begin{equation}
K_1\nabla\cdot(\nabla\theta-2e{\bf A})+K_{ij}\nabla_iB_j=0
\end{equation}
which is equivalent to the continuity equation for the current ($\nabla \cdot {\bf J}=0$).  The minimization with respect to ${\bf A}$ yields
\begin{equation}
{\bf J}_i= \frac{1}{4\pi}[\nabla\times ({\bf B}-4\pi {\bf M})]_i=-\frac{1}{4\pi \lambda^2}[{\bf A}_i-\frac{1}{2e}\nabla_i\theta-
\sum_j \sigma_{ji}{\bf B}_j]
\label{London}
\end{equation}
with \begin{equation}
4\pi {\bf M}_i=\frac{1}{\lambda^2}\sum_j\sigma_{ij}({\bf A}_j-\frac{1}{2e}\nabla_j\theta),
\label{mag}
\end{equation}
 $1/\lambda^2=8\pi (2e)^2K|\eta|^2$ and $\sigma_{ij}=16\pi e \lambda^2 K_{ij}$.
We take the surface normal is along the $\hat{z}$ direction and that the applied field is oriented along the $\hat{y}$ direction. Note that by applying an appropriate rotation to the fields in the free energy, this geometry results in no loss of generality. We assume that there are spatial variations only along the direction of the surface normal ($z$). We therefore have from $\nabla\cdot {\bf B}=0$ that $B_z=0$. We further choose ${\bf A}=[A_x(z),A_y(z),0]$ so that ${\bf B}=(-\partial A_y/\partial z,\partial A_x/\partial z,0)$ and work in a gauge where $\nabla\theta=0$. The three components of Eq.~\ref{London} yields
\begin{eqnarray}
\frac{\partial B_y}{\partial z}=&&\frac{1}{\lambda^2}\frac{\partial}{\partial z}[\sigma_{yy}A_y+\sigma_{zy}A_z]+\frac{1}{\lambda^2}A_x-\frac{1}{\lambda^2}\sigma_{xx}B_x \label{lon1}\\
\frac{\partial B_x}{\partial z}=&&\frac{1}{\lambda^2}\frac{\partial}{\partial z}[\sigma_{xx}A_x+\sigma_{zx}A_z]-\frac{1}{\lambda^2}A_y-\frac{1}{\lambda^2}\sigma_{yy}B_x \label{lon2}\\
4\pi J_z=&&0=A_z-\sigma_{zx}B_x-\sigma_{zy}B_y\label{lon3}.
\end{eqnarray}
Note that contributions from $\sigma_{xy}$ and $\sigma_{yx}$ cancel in the above. Taking derivatives of Eq.~\ref{lon1} and \ref{lon2} with respect to $z$, using Eq.~\ref{lon3} to eliminate $A_z$, we find
\begin{eqnarray}
(1-\frac{\sigma_{zy}^2}{\lambda^2})\frac{\partial^2 B_y}{\partial z^2}=&&\frac{1}{\lambda^2}B_y-\frac{\sigma_{xx}+\sigma_{yy}}{\lambda^2}\frac{\partial B_x}{\partial z}+\frac{\sigma_{zy}\sigma_{zx}}{\lambda^2}\frac{\partial^2 B_x}{\partial z^2}\\
(1-\frac{\sigma_{zx}^2}{\lambda^2})\frac{\partial^2 B_x}{\partial z^2}=&&\frac{1}{\lambda^2}B_x+\frac{\sigma_{xx}+\sigma_{yy}}{\lambda^2}\frac{\partial B_y}{\partial z}+\frac{\sigma_{zy}\sigma_{zx}}{\lambda^2}\frac{\partial^2 B_y}{\partial z^2}.
\end{eqnarray}
The above must be solved with the boundary conditions $B_i(z\rightarrow \infty)=0$ and
\begin{eqnarray}
H_y=&&B_y(z=0)-4\pi M_y(z=0)\\
0=&&B_x(z=0)-4\pi M_x(z=0)
\end{eqnarray}
where $H_y$ is the applied field. $M_x,M_y$ can be found using Eq.~\ref{mag} and Eq.~\ref{lon1}, \ref{lon2}, and \ref{lon3} to eliminate $A_x,A_y,$ and $A_z$ in favor of $B_x$ and $B_y$ and their derivatives. By setting $B_i=B_{i0}\exp(-\delta z/\lambda)$, the solution can be found analytically. The general form of the solution is quite involved, so here we present the solution for point groups $O$ and $C_{4v}$.

\subsubsection{$O$ point group}

A representative material is Li$_2$Pt$_3$B \cite{LiPt-PdB,yua06}. This problem has been solved in Refs.~\cite{LNE85,LuY08}. In this case there is only one Lifshitz invariant: $K_1{\bf B}\cdot{\bf j}$. Since this is a scalar under rotations the solution is the same for any orientation of the surface normal. The equations for ${\bf B}$ become:
\begin{eqnarray}
\frac{\partial^2 B_y}{\partial z^2}=&&\frac{1}{\lambda^2} B_y+ \frac{\delta}{\lambda^2}\frac{\partial B_x}{\partial z}\\
\frac{\partial^2 B_x}{\partial z^2}=&&\frac{1}{\lambda^2} B_x- \frac{\delta}{\lambda^2}\frac{\partial B_y}{\partial z}.
\end{eqnarray}
where $\delta=-2\sigma_{xx}$ (note $\sigma_{xx}=\sigma_{yy}$ in this case).
This coupled set of equations can solved for $B_{\pm}=B_x\pm iB_y$ \cite{LNE85,LuY08} with the result that to first order in $\delta/\lambda$:
\begin{eqnarray}
B_y=&&H_y\big[\cos\frac{\delta z}{\lambda^2}+\frac{\delta}{\lambda}\sin\frac{\delta z}{\lambda^2}\big]e^{-z/\lambda}\\
B_x=&&H_y\big[\frac{\delta}{\lambda}\cos\frac{\delta z}{\lambda^2}-\sin\frac{\delta z}{\lambda^2}\big]e^{-z/\lambda}.
\end{eqnarray}
Physically, this implies that the the magnitude of the $B_x$ is discontinuous as it crosses the surface (though not that of $B_y$) and that ${\bf B}$ also rotates inside the superconductor.  Note that in a slab geometry, $B_x$ is of opposite sign on the two sides of the slab. It may be possible to observe this through muon spin resonance experiments.

\subsubsection{$C_{4v}$ point group}

A representative material is CePt$_3$Si \cite{Bauer04}. In this case, the single Lifshitz invariant is generated by a Rashba spin-orbit coupling and is given by $K_1\hat{z}\cdot {\bf B}\times {\bf j}$. This implies $\sigma=\sigma_{xy}=-\sigma_{yx}\ne0$. The solution of the London problem now depends upon surface orientation and has been considered in Ref.~\cite{yip07}. We consider two situations here: the surface normal along and perpendicular to $\hat{z}$ (the four-fold symmetry axis). Consider first the normal along the $\hat{z}$ direction (in this case the applied field is $H_y$ and we find that $B_x=0$), then we have the usual London equation
\begin{equation}
\frac{\partial^2 B_y}{\partial z^2}=\frac{1}{\lambda^2}B_y\\
\end{equation}
with the unusual boundary condition $H_y|_{z=0}=(B_y+\frac{\sigma}{\lambda}B_y)|_{z=0}$. This yields the solution
\begin{equation}
B_y(z)=\frac{H_y}{1+\frac{\sigma}{\lambda}}e^{-z/\lambda}
\end{equation}
These equations show that there is no rotation of ${\bf B}$ across the sample surface. However, the magnetic induction ${\bf B}$ is discontinuous as the surface is crossed. Again, in a slab geometry, the discontinuity in $B_y$ is opposite for the two sides of the slab.

For the surface normal perpendicular to the $\hat{z}$ direction, the situation is different. To be concrete, consider the normal along the $\hat{x}$ direction and the applied field along the $\hat{y}$ direction (for the field along the $\hat{z}$ direction the usual London Equations result). In this case, it is again permissible to set $B_z=0$ and solve for $B_y$ to find
\begin{equation}
B_y=\frac{H_y}{1-\frac{\sigma^2}{\lambda^2}}e^{-z/\tilde{\lambda}}
\end{equation}
where $\sigma=\sigma_{xy}$ and $\tilde{\lambda}=\lambda(1-\frac{\sigma^2}{\lambda^2})$.

\subsection{Spatial structure of a single vortex}

\label{subsec:5}

The London theory can also be used to examine the field distribution of a vortex in a strongly type II superconductor. Again, the lack of inversion symmetry introduce some new physics. Here we focus (as above) on two examples with point groups $O$ and $C_{4v}$ and provide the solutions of Refs.~\cite{yip07,yip-cond-mat,LuY08,LuY09}. The approach used in these publications is to consider the parameter $\sigma_{ij}/\lambda$ to be small and then the Lifshitz invariants perturb the usual London solution. When there are no Lifshitz invariants, the solution to the London equations are $\theta=-\phi$ ($\phi$ is the polar angle) and the field is applied along the ${\hat n}$ direction
\begin{equation}
{\bf B}=\frac{1}{2e\lambda^2}K_0(r/\lambda) \hat{n}
\end{equation}
where $K_0(x)$ is a modified Bessel function. The perturbative solutions depend upon the specific form of the Lifshitz invariants and we turn to a discussion of two cases in turn.

\subsubsection{$O$ point group}

The solution in this case was found in Ref.~\cite{LuY08,LuY09}. The modified London equation is (the problem does not depend upon field direction)
\begin{equation}
\nabla\times \nabla \times {\bf A}+\frac{1}{\lambda^2}{\bf A}=\frac{\nabla \phi}{2e\lambda^2}+2\frac{\delta}{\lambda^2}\nabla \times {\bf A}-\frac{\pi\delta}{e\lambda^2}\delta^2({\bf r})\hat{z}.
\end{equation}
The new term implies that, in addition to the field along $\hat{z}$, there is an additional component along $\hat{\phi}$. The authors of Ref.~\cite{LuY08,LuY09} find that to first order in $\delta/\lambda$ the additional field is
\begin{eqnarray}
B_{\phi}^{(1)}(x=r/\lambda)=&\frac{\delta}{e \lambda^3}\Big\{K_1(x)\int_0^xx'dx'I_1(x')K_1(x')+I_1(x)\int_x^{\infty}x'dx'[K_1(x')]^2\Big\} \nonumber \\ &-  \frac{\delta}{2e\lambda^3}K_1(x)
\end{eqnarray}
where $I_1$ and $K_1$ are modified Bessel functions of the first kind.

\subsubsection{$C_{4v}$ point group}

The solution in this case was found in Ref.~\cite{yip07,yip-cond-mat}. The fields that appear due to the Lifschitz invariants depend in this case upon the orientation of the field. For the field along the $\hat{y}$ direction, it is found that the solution for ${\bf B}$ is given by (correct to first order in $\sigma/\lambda$) \cite{yip07,yip-cond-mat}
\begin{equation}
{\bf B}=\frac{1}{2e\lambda^2}K_0(|{\bf r}+\frac{\sigma}{\lambda}\hat{z}|/\lambda) \hat{y}.
\end{equation}
Physically, this implies that the maximum value of $B_y$ is shifted from the vortex center. This shift has also been seen in a full numerical solution of the Ginzburg Landau equations \cite{OIM06}. For the field along the $\hat{z}$ direction (the four-fold symmetry axis), the ${\bf B}$ field is unchanged and there is an induced magnetization along the radial direction \cite{yip07} (this radial magnetization was also found in the vortex lattice solution near $H_{c2}$ \cite{KAS05}).

\subsection{Vortex Lattice Solutions}

\label{subsec:6}
For fields near the upper critical field, there have been a variety of studies on the Abrikosov vortex lattice \cite{KAS05,hia08,mat08,hia09}. Some of these studies predict multiple phase transitions in the vortex lattice state \cite{hia08,mat08,hia09}. These studies are based on microscopic weak-coupling theories and involve an interplay of paramagnetism, orbital diamagnetism, gap symmetry, band structure, and spin-orbit coupling \cite{hia08,mat08,hia09}. While this chapter will not address these vortex lattice transitions, we will address some of the microscopic issues in the next chapter. Here we focuss on the GL theory, for which the predictions are more straightforward. In particular, near the upper critical field, the magnetic field is approximately uniform and the considerations above imply that the vortex lattice is hexagonal (perhaps distorted by uniaxial anisotropy). Consequently (following the arguments of Section II B) , the order parameter solution near the upper critical field is
$\eta({\bf r})=cnst \exp(i{\bf q}\cdot {\bf r}) \phi_0(x,y)$ where $\phi_0(x,y)$ is a lowest Landau level (LLL) solution. This solution, combining a phase factor and a (LLL) solution, has been called the helical vortex phase. The primary consequence of this solution is that the upper critical field is enhanced due to the presence of the Lifshitz invariants \cite{KAS05}. We note that due to the degeneracy of the LLL solution, there is ambiguity in the existence of the phase factor. In particular, the LLL solution ${\tilde \phi}_0(x,y)=e^{i\tau_yx/l_H^2}\phi_0(x,y-\tau_y)$ ($l_H$ is the magnetic length) is degenerate with $\phi_0(,x,y)$, consequently in some circumstances the wavevector ${\bf q}$ can be removed in favor of a shift of origin. This can be done whenever ${\bf q}$ is perpendicular to the applied magnetic field (this is the case for $C_{4v}$ point group symmetry but not for $O$ point group symmetry).  We feel that is still meaningful to speak of the helical vortex phase for the point group $C_{4v}$ because the same phase factor implies an increase of the in-plane critical field in two-dimensions for which this ambiguity does not exist. The name helical vortex phase reveals the link between the solutions in two and three dimensions.

In addition to studies near the upper critical field, there has been one numerical study of the time dependent GL equations in the vortex phase \cite{OIM06}. This study found the surprising result that the vortices flow spontaneously, in spite of the lack of an applied current. The claim is that the paramagnetic supercurrent (the magnetization current $\nabla\times {\bf M}$) is the origin of this spontaneous flux flow. We note that in this study the following boundary condition was used: ${\bf B}_{outside}={\bf B}_{inside}$. This differs from the continuity of ${\bf H}={\bf B}-4\pi {\bf M}$ discussed above. In the problem that was studied,  ${\bf M}$ is non-trivial and an examination of its neglect in the boundary condition can be seen to be equivalent to having a current flow. We argue that this current is cause the spontaneous flux flow. We note that the boundary conditions discussed here should be used in problems where the minimum length scale is $\xi_0$, the zero temperature coherence length. However, at lengths scale smaller than this, a microscopic theory is required and the single particle quantum mechanical wavefunctions will obey quite different boundary conditions.

\subsection{Multi-component order parameters}

\label{subsec:7}

There have not been as many studies on Lifshitz invariants in non-centrosymmetric superconductors in cases when the order parameter contains more than one complex degree of freedom. There has been one noteworthy result, which is the appearance of the helical phase when no magnetic fields are applied \cite{yua06,MinSam08}. In particular, if the ground state of the multi-component order parameter breaks time-reversal symmetry \cite{SU91,Book}, then the lack of both parity and time-reversal symmetries allows the helical phase to appear. As an example, consider the three dimensional irreducible representation of the point group $O$, with an order parameter  ${\vec \eta}$ where the components transform as the $(x,y,z)$ component of a vector.  The following Lifschitz invariant exists \cite{MinSam08}
\begin{equation}
iK(\eta_1^*D_y\eta_3+\eta_2^*D_z\eta_1+\eta_3^*D_x\eta_2 -c.c.).
\end{equation}
This Lifschitz invariant leads to a ground state order parameter ${\vec\eta}=e^{iqz}(1,i,0)$. The state ${\vec\eta}=(1,i,0)$ breaks time reversal symmetry and thus mimics the role of the magnetic field in the single component case.

\section{Microscopic Theory}

\label{sec:3}

The phenomenological arguments of the previous section have also been the subject of many microscopic calculations. These calculations, while all related, focus on and extend different aspects of the phenomenological theory above. In particular, four points of contact exist between the phenomenological theories and the microscopic theories. These are: direct calculations of the Lifshitz invariants in the free energy in Eq.~ \ref{glfree};  calculations of the magnetization in Eq.~\ref{mag1}; calculations of the current in Eq.~\ref{GLcurrent}; and calculations of the helical wavevector ${\bf q}$ in Eq.~\ref{hel}. We briefly review the first three of these and then turn to a more complete overview of microscopic studies of the helical phase since this turns out to be closely linked to FFLO phases.

\subsection{Contact between microscopic and macroscopic theories: Lifshitz Invariants}

\label{subsec:8}

The direct calculation of the Lifschitz invariants in Eq.~\ref{glfree} has been carried out by a few authors \cite{KAS05,Edel96,Y02,MinSam08} and can be found in Chapter 1 of this book.  In particular, the non-interacting Hamiltonian is
\begin{equation}
\label{H_0}
    H_0=\sum\limits_{{\bf k}}\sum_{\alpha\beta=\uparrow,\downarrow}
[\xi({\bf k})\delta_{\alpha\beta}+\mbox{\boldmath$\gamma$}({\bf k})
   \cdot \mbox{\boldmath$\sigma$}
 _{\alpha\beta}]
    a^\dagger_{{\bf k}\alpha}a_{{\bf k}\beta}
\end{equation}
where $ a^\dagger_{{\bf k}\alpha} $ ($a_{{\bf k}\alpha}$) creates (annihilates) an electronic state $ | {\bf k} \alpha \rangle $,  $\xi({\bf k})=\varepsilon({\bf k})-\mu$ denotes the spin-independent part of the spectrum measured relative to the chemical potential $ \mu$, $\alpha,\beta=\uparrow,\downarrow$ are spin indices,  $ \mbox{\boldmath$\sigma$}$ are the Pauli matrices, and the sum over ${\bf k}$
is restricted to the first Brillouin zone. In the helicity basis, this Hamiltonian is diagonalized with
energy bands given by
\begin{equation}
\xi_{\pm}({\bf k})=\xi({\bf k})\pm |\mbox{\boldmath$\gamma$}({\bf k})|
\end{equation}
with the Hamiltonian
\begin{equation}
\label{H_0 band}
    H_0=\sum_{{\bf k}}\sum_{\lambda=\pm}\xi_\lambda({\bf k})c^\dagger_{{\bf k}\lambda}c_{{\bf k}\lambda} ,
\end{equation}
where the two sets of electronic operators are connected by a unitary transformation,
\begin{equation}
a_{{\bf k}\alpha}=\sum_{\lambda}u_{\alpha\lambda}({\bf k})c_{{\bf k}\lambda},
\label{trans}
\end{equation}
with
\begin{equation}
\label{Rashba_spinors}
  ( u_{\uparrow\lambda}({\bf k}),~~ u_{\downarrow\lambda}({\bf k})) =
   \frac{( |\mbox{\boldmath$\gamma$}|+\lambda\gamma_z ,~~ \lambda (\gamma_x+i\gamma_y) )}{\sqrt{2|\mbox{\boldmath$\gamma$}|(|\mbox{\boldmath$\gamma$}|+\lambda\gamma_z)}}.
\end{equation}
In the limit that only one of the bands cross the the Fermi energy (this can be realized for superconductivity at the surface of a topological insulator \cite{san10}),
the following weak-coupling result for the coefficients defining the Lifishitz invariants of Eq.~\ref{glfree} is found
\begin{equation}
K_{ij}=-\frac{\mu_B N_0S_3}{2}\langle\phi^2({\bf k})
\hat{\bf{\gamma}}_i({\bf k})v_j({\bf k})\rangle \label{lif-1}
\end{equation}
where $N_0$ is the density of states of the band at the chemical potential, $\phi({\bf k})$ describes the superconducting state and is an even function belonging to one of one-dimensional representations
of the point group of the crystal, $\langle...\rangle$ means the averaging over the Fermi surface, $\mu_B$ is the Bohr magneton,  and
\begin{equation}
S_3(T)=\pi  T\sum_n\frac{1}{|\omega_n|^3}=\frac{7\zeta (3) }{4\pi^2T^2}.
\label{S_3}
\end{equation}

Eq.~\ref{lif-1} is valid when there is only a single band present.  When two bands are present (as is often the case), and assuming that $\phi({\bf k})$ is the same for both bands, then Eq.~\ref{lif-1} must be multiplied by the factor
\begin{equation}
\delta N = (N_+-N_-)/(N_++N_-).
\end{equation}
where $N_{\pm}$ are the density of states of the two bands ($N_0=N_++N_-$). Microscopic calculations of the Lifshitz invariants are limited to the regime near $T_c$ where the GL theory is valid.

\subsection{Contact between microscopic and macroscopic theories: current and magnetization}

\label{subsec:9}

In the limit of small magnetic fields (${\bf B}$) and small phase gradients ($\nabla \theta$) in the superconducting order parameter, it it possible to find microscopic extensions to Eq.~\ref{GLcurrent} and Eq.~\ref{mag1} that are valid for all temperatures. This has been carried out in Refs.\cite{Y02,Edel95}. Here, we follow the notation of Ref.~\cite{Y02}. In the clean limit, for 2D cylindrical bands with a Rashba interaction ($\mbox{\boldmath$\gamma$}({\bf k})=\alpha \hat{n}\times {\bf p}({\bf k})$) Eq.~\ref{GLcurrent} and Eq.~\ref{mag1} can be rewritten as
\begin{eqnarray}
J_x=&\rho_s\frac{\hbar \nabla_x \theta}{2m}-\kappa B_y \nonumber\\
M_y=&\frac{\kappa}{2} \hbar \nabla_x \theta
\end{eqnarray}
where $M_y$ is the magnetic moment, $\rho_s$ is the superfluid density, and
\begin{equation}
\kappa(T)=\frac{\mu}{4\pi \hbar^2}[p_{F+}\{1-Y(T,\Delta_+\}-p_{F-}\{1-Y(T,\Delta_-)\}] \label{kap}
\end{equation}
where $p_{F,\pm}$ are the Fermi momenta for the two bands, $\Delta_{\pm}$ are the gaps on the two bands, $\mu$ is the Fermi energy, and $Y(x)$ is the Yoshida function. Note that Eq.~\ref{kap} is  proportional to $\delta N$ in the limit $\delta N <<1$.

The role of Fermi liquid corrections has also been examined \cite{Fuj05} in this context. This study has found the that the only Fermi liquid corrections that alter the current contribution from the Lifshitz invariants are ferromagnetic correlations. If there are no ferromagentic correlations, then Eq.~\ref{kap} is unchanged. This is important in heavy Fermion materials, where the effective mass enhancement suppresses the usual supercurrent but does not change Eq.~\ref{kap} \cite{Fuj05}.

\subsection{Microscopic Theory of the Helical and FFLO Phases}

\label{subsec:10}

The helical phase has received a great deal of attention from the microscopic point of view
\cite{BG02,KAS05,Sam08,hia09,mat08,hia09,DF03,dim07,agt07,san10}. One reason for this is that it is closely related to the FFLO phase \cite{ff,lo} in which the superconducting order parameter develops a periodic spatial structure. The interplay between these two phases is not trivial. It is perhaps not surprising that spatially oscillating superconductor solutions readily appear in non-centrosymmetric superconductors when magnetic fields are applied. In particular, a state with
momentum ${\bf k}$ at the Fermi surface will generally not have a
degenerate partner at $ - {\bf k} $ with which to form a Cooper pair when both parity and time reversal symmetries are broken.
The state ${\bf k} $ would rather pair with a degenerate state $ -{\bf k}
+ {\bf q}$ and in this way generate a spatially oscillating superconducting
order parameter.

\begin{figure}
\sidecaption
\includegraphics[scale=.65]{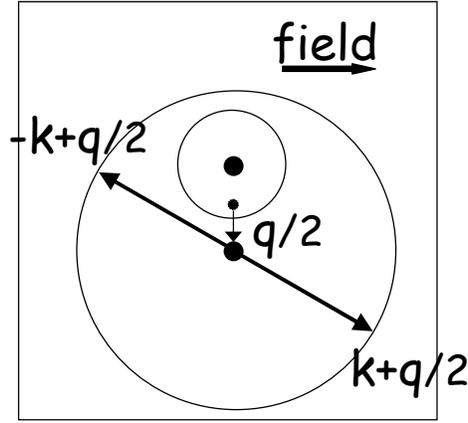}
\caption{A
magnetic field directed as shown together with a Rashba spin-orbit interaction shifts the center of the large and small
Fermi surfaces by $\pm{\bf q}/2$. The smaller dot represent the point
$(0,0)$ (center of Fermi surfaces without field) and the two
larger dots represent the points $(0,-{\bf q}/2)$ and $(0,{\bf q}/2)$
(centers of the new Fermi surfaces). Pairing occurs between states of ${\bf k}+{\bf q}/2$
and $-{\bf k}+{\bf q}/2$, leading to a gap function that has a spatial
variation $\Delta({\bf x})=\Delta_0\exp(i{\bf q}\cdot{\bf x})$. From Ref.~\cite{agt07}.} \label{fig1}
\end{figure}
\vglue 0.5 cm

The microscopic origin of the spatially oscillating
states can be understood by an examination of the single particle
eigenstates when a Zeeman field ${\bf H}$ is included (for now we ignore the vector potential ${\bf A}$)
\begin{equation} H_Z=-\sum_{{\bf k},\alpha,\beta}\mu_B{\bf H}\cdot \mbox{\boldmath$\sigma$} _{\alpha\beta}
    a^\dagger_{{\bf k}\alpha}a_{{\bf k}\beta}.\end{equation}  The single particle
excitations now become
\begin{equation}
    \xi_{\pm}({\bf k},{\bf H})=\xi({\bf k})\pm \sqrt{\mbox{\boldmath$\gamma$}^2({\bf k})-2\mu_B\mbox{\boldmath$\gamma$}({\bf k})\cdot {\bf H}+
   \mu_B^2{\bf H}^2}.
\label{e3}
\end{equation}
In the limit $|\mbox{\boldmath$\gamma$}|>>|H|$, this becomes (we ignore the small
regions of phase space for which $\mbox{\boldmath$\gamma$}=0$)
\begin{equation}
\xi_{\pm}({\bf k},{\bf H}) \approx \xi({\bf k})
\pm\mu_B\hat{\mbox{\boldmath$\gamma$}}({\bf k})\cdot{\bf H}.
\end{equation}
The origin of pairing states with non-zero ${\bf q}$ (that is $\Delta({\bf x})\propto e^{i{\bf q}\cdot {\bf x}}$) follow from this expression.  As an example, consider a Rashba interaction
$\mbox{\boldmath$\gamma$}=\gamma_{\perp}(k_y\hat{x}-k_x\hat{y})$ for a cylindrical Fermi surface and
a magnetic field along $\hat{x}$. In this
case, as shown in Fig.~\ref{fig1}, the Fermi surfaces remain circular and the centers are
shifted along the $\hat{y}$ direction.  A finite center of mass
momentum Cooper pair is stable because the same momentum vector
${\bf q}$ can be used to pair {\it every} state on one of the two
Fermi surfaces. In the more general case, for a non-zero ${\bf q}$
state to be stable, the paired states should be degenerate:
$\xi_{\pm}({\bf k}+{\bf q},{\bf H})=\xi_{\pm}(-{\bf k}+{\bf q},{\bf H})$, this gives the condition $\hbar
{\bf q}\cdot {\bf v}_F=\mu_B{\bf H}\cdot\hat{\mbox{\boldmath$\gamma$}}({\bf k})$. This differs from the condition for
the usual FFLO phase, for which $\hbar {\bf q}\cdot {\bf v}_F=\mu_B|{\bf H}|$. The
optimal paring state corresponds to finding ${\bf q}$ that satisfies
the pairing condition for the largest possible region on the Fermi surface.

\begin{figure}
\includegraphics[scale=.6]{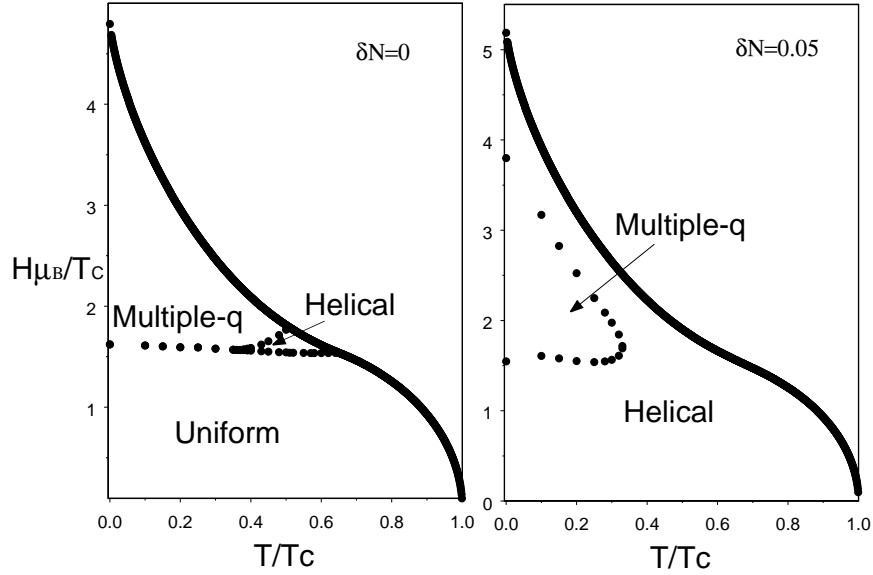}
\caption{Typical phase diagram showing both multiple-${\bf q}$ and single-${\bf q}$ (helical phase) phases as a function of Zeeman field in a clean non-centrosymmetric superconductor for two different values of $\delta N$. These calculations where carried out with a Rashba spin-orbit interaction and a 3D spherical Fermi surface (a 2D cylindrical Fermi surface gives similar results). For fields $H\mu_B/T_c <1.5$, $q\approx \delta N H\mu_B/ v_F$ (for the $\delta N=0$, this leads to ${\bf q}=0$),  while for higher fields $q\approx  H\mu_B/ v_F$. From Ref.~\cite{agt07}.} \label{fig2}
\end{figure}
\vglue 0.5 cm

The above paragraph also reveals the origin of the interplay between the helical and FFLO phases. In particular, the two Fermi surface sheets prefer pairing states with opposite sign of ${\bf q}$. Choosing a particular ${\bf q}$ allows pairing on one Fermi surface, but not on the other. This naturally leads to competition between single-${\bf q}$ (helical) and multiple-${\bf q}$ (FFLO-like) states. Which state appears depends upon the details of the system. Without going into further microscopic details, which can be found in Refs.~\cite{BG02,KAS05,Sam08,hia09,mat08,DF03,dim07,agt07,san10}, we summarize some of the main results here. One important result  is that since there are two sources of the modulation ${\bf q}$ (FFLO-like physics and Lifshitz invariants), there are two typical values for the magnitude of $q$ \cite{dim07,agt07,Sam08,hia09} that both appear in different regions of the temperature/magnetic field phase diagram. In particular $q\approx H\mu_B/ v_F$ stems from FFLO-like physics related to Fig.~\ref{fig1} and is the value of $q$ in the high-field regime (in clean materials). While $q\approx \delta N H\mu_B/ v_F$ stems from the Lifschitz invariants and is the typical magnitude of $q$ in the low-field regime \cite{dim07,agt07,Sam08}. As shown in Fig.~\ref{fig2}, in the clean limit, both single-${\bf q}$ and multiple-${\bf q}$  phases exist \cite{agt07,dim07}. However, the multiple-${\bf q}$ phase become less stable as $\delta N$ increases \cite{agt07}.  We note that in the case of superconductivity at the surface of a topological insulator, which is akin to $\delta N=1$, only the single-${\bf q}$ exists \cite{san10}. In the dirty limit the multiple-${\bf q}$ phases no longer appear, while the single-${\bf q}$ phase with $q\approx \delta N H\mu_B/ v_F$ is robust \cite{Sam08,dim07}. Finally we note that when the vector potential is also included then novel vortices and vortex phases may appear \cite{dim07,agt08,mat08,hia08,hia09}.

\section{Conclusions}

\label{sec:4}

In this chapter we have examined the role of Lifshitz invariants that appear in the Ginzburg Landau free energy of non-centrosymmetric superconductors. These invariants lead to magnetoelectric effects, novel London physics in the Meissner state, new structure in individual vortices, and a helical phase in which the order parameter develops a periodic spatial variation. Additionally, we have provided an overview of theoretical developments in the microscopic description of this physics.

\begin{acknowledgement}
The author would like to thank S. Fujimoto, K. Samokhin, and M. Sigrist for useful discussions. This work was supported by NSF grant DMR-0906655.
\end{acknowledgement}

%
%

%
%
%

\end{document}